\documentclass{iau}
\usepackage{graphicx}

\title[The kHz QPOs  of  Neutron Stars and Millisecond Pulsars in LMXBs] 
{The kHz QPOs  of  Neutron Stars and Millisecond Pulsars and
Implications}

\author[Chengmin  Zhang and Dehua Wang]   
{Chengmin  Zhang and Dehua Wang
}

\affiliation{National Astronomical Observatories, Chinese Academy of
Sciences, Beijing 100012,   China }

\pubyear{2012}
\volume{290}  
\pagerange{119--126}
\jname{Feeding compact objects: Accretion on all scales}
\editors{C.M. Zhang, T. Belloni, M. M\'endez \& S.N. Zhang, eds.}
\begin{document}

\maketitle

\begin{abstract}

The kilohertz quasi-periodic oscillations (kHz QPOs) have been found
in neutron star low mass X-ray binaries (NS-LMXBs), which present
the millisecond timing phenomena close to the surface of the compact
objects. We briefly  summarize the following contents: (1).  The
correlations and distributions of twin kHz QPOs; (2). The relations
of  high-low frequency QPOs; (3). The QPO properties  of NS Atoll
and Z sources;  (4). No clear direct correlations between NS spins
and QPOs;  (5). The mechanisms of kHz QPOs; (6). The implications of
 kHZ QPOs, e.g., NS mass and radius,  disk  thickness
 and magnetic field of Atoll and Z source.

\end{abstract}

\firstsection 

\section{Introduction}
The kHz QPOs in NS-LMXBs were firstly discovered in 1996 in both
Atoll and Z sources (Hasinger and van der Klis 1989)
since the launch of RXTE in the end of 1995
(\cite[van der Klis 2000,2006]{van der Klis00}).  Although RXTE has
finished its destination journey, the data of RXTE are still coming
out for  analysis.   The QPO have been found in about 40 NS-LMXBs,
where 28 sources are found the twin kHz QPOs, which are believed to
be the millisecond oscillations around the surface of compact
objects, thus kHz QPO observations   provide us  a powerful tool to
probe the strong gravity regimes of  compact objects (\cite[van der
Klis 2006,2008]{van der Klis06}). Through kHz QPO explorations, we
could know the properties of Keplerian motion or Einstein gravity
near compact objects, and infer  the mass, radius and magnetic field
of NS, and the accretion flow and magnetosphere of NS. We can
understand the details of formation process of millisecond pulsar
while NS is recycled in the accretion system (Alpar et al. 1982;
Chakrabarty \& Morgan 1998).  Furthermore, kHz QPOs will help us the
general properties of accretion physics in binary compact  systems
(Bhattacharya \& van den Heuvel 1991; Tauris et al. 2012; Liu Q.Z.
et al. 2007; van den Heuvel 2004; Shore et al. 2010).

\section{Brief Summary of kHz QPOs}
{\bf Correlations and  distributions  of twin kHz QPOs} Now, it is
clear  that the frequencies of twin kHz QPOs follow a non-linear
relation, or in other words the constant ratio or beat with the spin
frequency for the twin QPOs cannot work ( \cite[Belloni et al.
2007]{Belloni07}).  Twin kHz QPOs  (upper $\nu_2$ and lower $\nu_1$)
show a power law relation (\cite[Zhang et al. 2006]{Zhang06}).   The
linear correlation between  HBO and lower kHz QPO has been reported,
which can be extended to black hole or even white dwarf binary
systems (Belloni  et al. 2002, 2005, Psaltis et al. 1999).
%
Like Cir X-1,  all QPO data seem to follow a parabola route in the
plot of  difference of twin QPOs ($\Delta\nu$ = $\nu_2$ -  $\nu_1$ )
vs. upper frequency ($\nu_2$). However, for the individual Atoll or
Z source, the kHz QPOs are usually detected in the frequency regime
of high frequency, e.g. from 500 Hz - 1200 Hz. This correlation
leave a strong constrain on the QPO models. At present, only two
models can predict this parabola trend,  the relativistic precession
model (RPM) by Stella and Vietri (1999) and Alfven wave oscillation
model (AWOM) by Zhang (2004). In RPM, the free parameter of model is
NS mass,  which are often overestimated by kHz QPO data to be about
2 solar masses, a bigger value than the averaged value (1.6 solar mass) of
millisecond pulsars (MSPs) (Zhang et al. 2011). In AWOM model, the
free parameter is the mass density of star, which is consistent with
the standard value of NS with 1.4 solar mass and 15 km.

{\bf The distinction of  QPO and spin frequency of Atoll and Z
sources.} Averagely, the maximum kHz QPO frequency of Atoll is a
slight bigger than that of Z sources, and the reason for this may be
the thickness of accretion disk, e.g. the disk of Z source should be
thicker than that of Atoll, which would be also the reason why the
spin frequencies have not yet been detected in Z sources. Moreover,
the size of corona of Z source would be bigger than that Atoll,
where a lot of emission photons should be blocked there.

{\bf Q-factors of twin kHz QPOs.}  The Q-factor properties of Atoll
and Z sources are complicated \cite[Barret et al. 2011]{Barret11},
\cite[M\'endez 2006]{Mendez06}). Our statistics seems to infer a
rough correlation for the Q-factors of upper and lower kHz QPOs to
the accretion rate (Wang et al. 2011), and the implied mechanisms
for upper and lower should be different.  The smaller  averaged
Q-factor of Z sources than that of Atolls should be ascribed to the
thicker disks of Z sources (or corona size), and the quantitative
work on this point is going on.  The abrupt changing  of Q-factors
at some frequency seems to concorde to the changing of twin kHz QPO
correlation, however that the physics behind is ascribed to ISCO or
particular radius  has not yet been determined clearly.




{\bf The mechanisms of kHz QPOs} It has been long time debates on
the mechanisms of kHz QPOs.  Abramowicz et al (2003) wish to
consider the relativistic disk oscillations for QPO productions at
preferred radius, which can be applied to 3:2 QPO ratios of BHs, but
not for NS QPO ratio that has no a constant ratio of 3:2.   The
simple beat model cannot fit for the twin  kHz QPO data with the
detected spin frequency (Miller et al. 1998).   The relativistic
precession model (RPM) by Stella and Vietri (1999) ascribes the
upper to Keplerian orbital frequency of disk matter and the lower to
the relativistic perihelion precession frequency of accreting
matter. By exploiting this model, we compare the twin kHz QPOs of
Sax J 1808.4-3658 (Wijnands et al. 2003) with the model
prediction, and prefer the NS mass of this source is about 3.2 solar
mass,  a upper limit of NS mass before collapsing to BH, which is
much bigger than its estimated mass of 1.4 solar mass. Thus, RPM
should overcome the overestimation of NS mass.
 In Alfven wave oscillation model (AWOM, Zhang 2004, Zhang et al. 2007b, 2009),
the  upper is also considered as the  Keplerian orbital frequency,
and the lower is ascribed to the Alfven wave Oscillation frequency
of accreting matter along the orbit with magnetic field. We once
again apply AWOM model to Sax J 1808.4-3658, and find that its NS
radius is about 20 km for the mass of 1.4 solar mass. NS radius of
20 km is bigger than what we often presumed 15 km, then this value
has no contradiction to the predictions of nuclear physics or from
other arguments (Lattimer \& Prakash 2004; Menezes et al.
2004; Haensel et al.  2007).

{\bf The implications of  NS mass and radius,   NS  magnetic field}
NS mass and radius can be inferred by the highest  kHz QPO
frequency, by considering it as a Keplerian frequency  at NS surface
or ISCO.  The QPO models, e.g. RPM or AWOM, can also prefer NS mass
or its mass density ($M/R^{3}$), based on which we can compare the EOS
of nuclear matter (Miller 2002; Ozel 2006; Haensel 2007, 2008).
 Although the luminosity values
of Atoll and Z are very different, perhaps spanning 2 magnitudes,
then the   similar  kHz QPO distributions of them inferred that both
sources share the similar magnetosphere, which infers a 10 times
stronger field of Z source (Eddington luminosity) than that of Atoll
(0.01 Eddington luminosity).   To interpret the Atoll and Z changing
in Cir X-1 and XTE 1701,   our  arguments are below:   while the
luminosity is strong (weak),  then the source shows a Z state
(Atoll), since a strong luminosity (high accretion rate) will
correspond to a stronger field (inward the magnetosphere).

\section{Perspective of kHz QPOs}

The maximum QPO frequency may show the surface information of NS, by
which we can com=nstrain the NS radius and mass.  The 1200 Hz often
gives  the loose constrains of NS parameters, then this frequency
may occur at some preferred radius or barrier of magnetosphere-disk
before the accreting matter colliding the NS surface or ISCO. 1860
Hz has been reported, but such a high frequency may be a harmonic of
900 Hz (Bhattcharyya 2011).  The  area of deep  inside
magnetosphere-disk may absorbs and blurs  the emission QPO photons,
which can destroy or weaken the QPO signal below our detection
limit.  In any cases, we can employ the maximum QPO frequency to
explore the inner structure of accretion disk.  The HBO mechanism is
not clear. If it is considered as a similar mechanism to the lower
kHz QPO but occurred  at a far radius, or outer boundary of disk,
then  it should be 50 km for a kHz QPO production at 20 km. However,
the origin of  outer disk radii need more through study.  The spin
and kHz QPOs has no direct correlations, then the rough indirect
correlations may be possible, since averaged  kHz QPO frequency
seems to be proportionally related to spin frequency, which needs
the further quantitative statistics.

\noindent Supported by National Basic Research Program
of China  (2009CB824800, 2012CB821800),  National Natural Science
Foundation of China NSFC (11173034).

%


\end{document}